\documentclass{article}
\usepackage{dcase2025_techrep,amsmath,graphicx,url,times,booktabs, tabularx}
\usepackage{amssymb}
\usepackage[hidelinks]{hyperref}

\newcolumntype{C}{>{\centering\arraybackslash}X}
\newcolumntype{R}{>{\raggedleft\arraybackslash}X}
\newcolumntype{L}{>{\raggedright\arraybackslash}X}

\newcommand{\Fone}{$\text{F}_{20^\circ/1}$}
\newcommand{\LE}{$\text{LE}_\text{CD}$}

\newcommand{\RDE}{$\text{RDE}_\text{CD}$}

\newcommand{\ESELD}{$\mathcal{E}_{\text{SELD}}$}

\title{Improving Stereo 3D Sound Event Localization and Detection: Perceptual Features, Stereo-specific Data Augmentation, and Distance Normalization}

\name{Jun-Wei Yeow$\sthanks{This research is supported by the Ministry of Education, Singapore, under its Academic Research Fund Tier 2 (MOE-T2EP20221-0014)}$,
      Ee-Leng Tan,
      Santi Peksi,
      Woon-Seng Gan 
      }
\address{Smart Nation TRANS Lab, Nanyang Technological University, Singapore \\
junwei004@e.ntu.edu.sg, \{etanel, speksi, ewsgan\}@ntu.edu.sg\\          
 }

\begin{document}

\ninept
\maketitle

\begin{sloppy}

\begin{abstract}
This technical report presents our submission to Task 3 of the DCASE 2025 Challenge: Stereo Sound Event Localization and Detection (SELD) in Regular Video Content. We address the audio-only task in this report and introduce several key contributions. First, we design perceptually-motivated input features that improve event detection, sound source localization, and distance estimation. Second, we adapt augmentation strategies specifically for the intricacies of stereo audio, including channel swapping and time-frequency masking. We also incorporate the recently proposed FilterAugment technique that has yet to be explored for SELD work. Lastly, we apply a distance normalization approach during training to stabilize regression targets. Experiments on the stereo STARSS23 dataset demonstrate consistent performance gains across all SELD metrics. Code to replicate our work is available in this repository\footnote{\url{https://github.com/itsjunwei/NTU_SNTL_Task3}}

\end{abstract}


\section{Introduction}
\label{sec:intro}

Sound Event Localization and Detection (SELD) is a form of machine-listening that enables systems to not only understand what sounds are happening, but also where they come from~\cite{adavanne2018sound}. This form of spatial intelligence can be extended into three-dimensions by integrating Sound Distance Estimation (SDE), cumulating in 3D SELD. The transition of the Detection and Classification of Acoustic Scenes and Events (DCASE) challenge tasks from SELD to 3D SELD also signifies the growing interest in distance-aware systems~\cite{Diaz-Guerra2024_DCASE2024Task3}.

Most existing SELD and 3D SELD systems use small microphone arrays, often configured in either the First Order Ambisonics (FOA) or Multi-channel Microphone Array (MIC) formats. In 2025, the DCASE Challenge Task 3 shifts the focus onto stereo-based 3D SELD, pivoting to a form of spatial awareness meant for consumer electronics. This reflects a greater trend towards more consumer-friendly environmental intelligence, such as for wearables~\cite{nagatomo2022wearable} and online inference systems~\cite{yeow2024real}.

Compared to traditional FOA or MIC audio formats, stereo-based SELD has not yet been extensively explored~\cite{Wilkins2023_TwovsFour}. In this paper, we outline our proposed system and general methodology for stereo-based 3D SELD. Our methods, including stereo-aware augmentation, perceptually-inspired features, and distance normalization, can apply to generic stereo 3D SELD pipelines to significantly improve performance.

\section{Input Features}

Let $x_\mathrm{L}[n]$ and $x_\mathrm{R}[n]$ denote the left and right stereo input channels, respectively, with $n$ being the discrete-time index. The Short-Time Fourier Transform (STFT) of the $c$-th channel at time frame~$t$ and frequency bin~$f$ is denoted as $X_{c}(t,f)$, for $c \in \{\mathrm L, \mathrm R\}$. 

\subsection{Mid-Side Conversion}

Mid-Side (MS) conversion explicitly decomposes the stereo signal into Mid (M) and Side (S) components. This decomposition has been explored for acoustic analysis tasks using stereo audio, such as Acoustic Scene Classification~\cite{han2017convolutional}. The conversion process is performed in the time-domain as follows,

\begin{equation}
m[n] = \frac{x_{\mathrm{L}}[n] + x_{\mathrm{R}}[n]}{2}, \quad
s[n] = \frac{x_{\mathrm{L}}[n] - x_{\mathrm{R}}[n]}{2},
\label{eq:mid_side}
\end{equation}

\noindent where $m[n]$ and $s[n]$ denote the discrete-time mid and side signals, respectively. Here, $m[n]$ represents the average pressure and $s[n]$ captures the horizontal pressure differential. The STFTs of $m[n]$ and $s[n]$ are therefore $M(t,f)$ and $S(t,f)$, respectively. 

Similar to the Intensity Vector (IV) used in FOA-based SELD work~\cite{cao2019two}, we derive a MS-based intensity feature for stereo audio. For each time-frequency (TF) bin, the real portion of the MS cross-spectrum is computed as follows, 

\begin{equation}
  I_{x}(t,f) \;=\; \Re \bigl\{ M(t,f) \; S^{*}(t,f) \bigr\},
  \label{eq:intensity_ms}
\end{equation}

\noindent before being normalized by the total MS power: 

\begin{equation}
\label{eq:norm_Ivs}
    \Tilde{I}_{x}(t,f) \;=\; \frac{I_{x}(t,f)}{|M(t,f)|^2 + |S(t,f)|^2 + \varepsilon} \;. 
\end{equation}

\noindent where $\varepsilon$ is a small constant to prevent division by zero. Finally, $\Tilde{I}_{x}(t,f)$ is typically projected onto a $K$-band Mel scale using the Mel filter bank matrix $\mathbf{W}_{\mathrm{mel}}$:

\begin{equation}
    \label{eq:mel_IV}
    \mathrm{IV}(t,f) = \Tilde{I}_{x}(t,f) \cdot \mathbf{W}_{\mathrm{mel}}(f,k).
\end{equation}

This IV feature captures stereo intensity differences in a similar fashion to spatial features for FOA audio~\cite{perotin2019crnn_ivs}, providing important directionality cues to our stereo-based SELD system.

\subsection{Spatial Coherence}

Spatial coherence measures the similarity between channels as a function of frequency. This property has been investigated in many SDE frameworks due to its strong relationship with sound source distance~\cite{zhagyparova2021supervised}. In this work, we use the magnitude-squared coherence (MSC) between the LR channels.

Firstly, we define the cross-power spectral density between the two stereo channels as

\begin{equation}
\label{eqn:psd_expectation}
    \Phi_{\mathrm{L,R}}(t,f) = 
    \mathbb{E}
    \Bigl[
    X_\mathrm{L}(t,f) \, X^*_\mathrm{R}(t,f)
    \Bigr].
\end{equation}

In practice, we estimate $\Phi_{\mathrm{L,R}}(t,f)$ using time-recursive averaging~\cite{schwarz2015coherent}:

\begin{equation}
\label{eqn:recusrive_psd}
    \hat{\Phi}_{\mathrm{L,R}}(t,f) = 
    \lambda \, \hat{\Phi}_{\mathrm{L,R}}(t-1, f) 
    + (1-\lambda) \, X_\mathrm{L}(t,f) \, X_\mathrm{R}^*(t,f),
\end{equation}

\noindent where $\lambda \in [0,1]$ is a smoothing coefficient, set as $0.8$ in this work~\cite{jeub2011blind}. The MSC $\hat{\gamma}(t,f)$ is subsequently calculated as

\begin{equation}
\label{eqn:gamma_ij}
    \hat{\gamma}(t,f) = 
    \frac{|\hat{\Phi}_{\mathrm{L,R}}(t,f)|^2}
    {\hat{\Phi}_{\mathrm{L,L}}(t,f) \, \hat{\Phi}_{\mathrm{R,R}}(t,f) + \varepsilon},
\end{equation}

\noindent where $0\!\leq\!\hat{\gamma}(t,f)\!\leq\!1$. Here, high MSC values typically signal direct, coherent sources (near/focused events), while lower values suggest diffuse or distant sources. Similarly, we project the MSC onto the same $K$-band Mel scale using $\mathbf{W}_\mathrm{mel}$:

\begin{equation}
    \label{eqn:Mel_MSC}
    \mathrm{MSC}(t,f) = \hat{\gamma}(t,f) \cdot \mathbf{W}_\mathrm{mel}(f,k).
\end{equation}

\section{Data Augmentation}

We employ both waveform-level and spectrogram-level augmentation methods to generate meaningful variations in stereo spatial cues, thereby improving model robustness.

\subsection{Waveform-level}

Audio channel swapping (ACS) methods have been developed for both the FOA and MIC audio formats~\cite{mazzon2019first, wang2023four}. ACS-based methods are extremely effective for the two-dimensional SELD task due to them being able to significantly increase the number of directional events, while preserving the natural reverberation conditions of the recording environments~\cite{niu2023experimental}. 

In the case of stereo audio, the ACS method becomes a simple swapping of the left and right channels. Accordingly, the azimuth labels are also inverted about the frontal axis. This can essentially double the amount of directional sound events available.

\subsection{Spectrogram-level}

In this work, we explore three different spectrogram-level data augmentation methods that are applied to the input features on-the-fly during training.

\textbf{FilterAugment} applies band-specific gains across the input spectrograms, simulating realistic distortions across frequency bands. Originally developed for Sound Event Detection~\cite{nam2022filteraugment}, this method introduces variability in spectral coloration. Therefore, this can prevent models from relying on frequency-specific artifacts, making it attractive and applicable for SELD. 

\textbf{Frequency Shifting} perturbs the input spectrograms by shifting frequencies within a controlled range, simulating pitch variation in the frequency domain. This has shown to improve generalization by encouraging the model to learn frequency-invariant representations of spatial cues~\cite{Nguyen2022_SALSALite, Nguyen2022_SALSA}.

\textbf{Inter-Channel-Aware Time-Frequency Masking (ITFM)} is our proposed adaptation of TF masking (TFM) for stereo audio. Traditional TFM methods, such as SpecAugment~\cite{park2019specaugment} or Cutout~\cite{devries2017improved_cutout}, typically apply independent or identical masks to each of the spectral channels. This risks distorting or erasing inter-channel differences, which are critical for robust localization and distance estimation in stereo-based tasks. Our ITFM method pre-computes and reapplies inter-channel differences post-masking. Therefore, this helps to preserve inter-channel differences, maintaining spatial information essential for robust 3D SELD.

\section{Network Architecture}

\begin{figure}[t]
    \centering
    \includegraphics[width=0.95\linewidth]{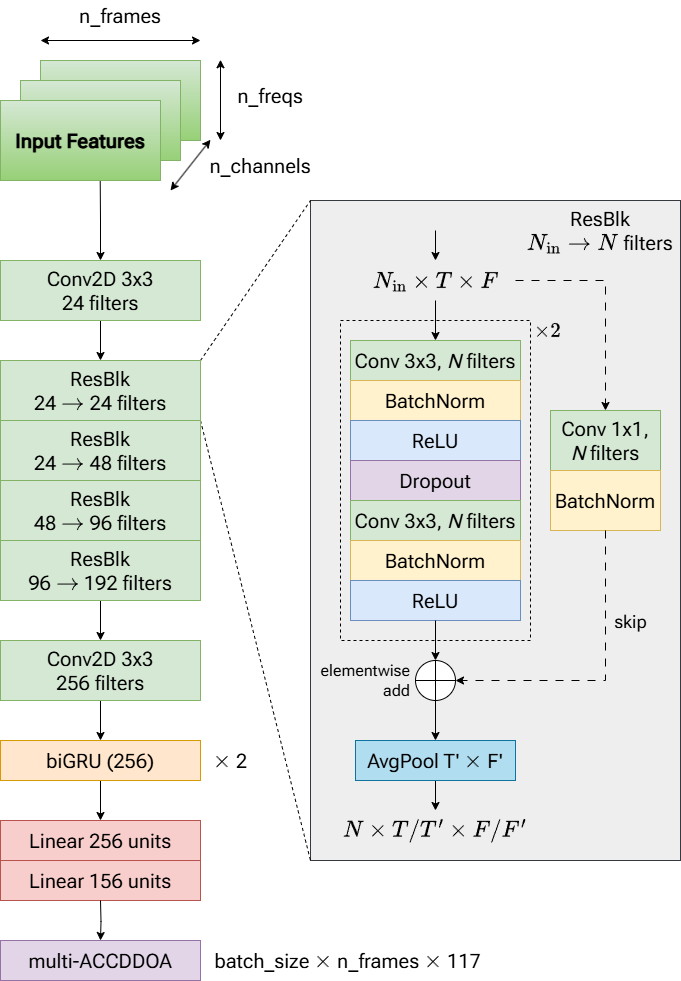}
    \caption{Block diagram of the ResNet-biGRU CRNN used in our DCASE 2025 submission.}
    \label{fig:SELDNet}
\end{figure}

For this year's challenge, we employed a relatively lightweight convolutional recurrent neural network (CRNN) built on a ResNet backbone followed by bi-directional gated recurrent units (biGRUs). The architecture follows similarly to our previous DCASE submission in 2024~\cite{Yeow_NTU_task3a_report}. For the output, we use the multi-ACCDDOA output format proposed by Shimada et al.~\cite{shimada_multiaccdoa} and extended for distance estimation by Krause et al.~\cite{krause2024sound}.

Figure~\ref{fig:SELDNet} showcases the CRNN system that we use for our submissions. The Mean Squared Error (MSE) is used as the loss function. Notably, we opt not to use Conformer modules~\cite{wang2023four, niu2023experimental} to reduce the substantial computational and environmental costs associated with training such large, complex models. Quantitatively, our full systems only uses around 4 million parameters, and requires 1.89G multiply-accumulate operations (MACs) per forward pass.

\section{Distance Normalization}
\label{sect:Distance_Normalization}

The distance values in the STARSS23 dataset can range from $[0.04, 7.64]$ in meters. If we were to directly regress these values in ACCDOA-based output format variants, the MSE loss function can very easily be biased towards further or more distant sound events~\cite{krause2024sound}. Therefore, to mitigate this problem, we apply the distance normalization method that was first proposed in our previous work on 3D SELD~\cite{Yeow_NTU_task3a_report}.

This distance normalization procedure scales the distribution of distances, $d$, to a uniform range of $[-1, 1]$ in two steps:

\begin{equation}
\label{eqn:dnorm}
    d' = \frac{d - \Bar{d}}{\sigma_d}, \quad
    d_\text{norm} = \frac{d'}{\max(d')},
\end{equation}

\noindent where $\Bar{d}$ and $\sigma_d$ represent the mean and standard deviation of all distances, respectively. This normalization ensures that all elements in the multi-ACCDDOA vector lies within the same scale of $[-1,1]$, preventing larger distances from disproportionately affecting the MSE loss, thereby yielding better overall 3D SELD performance.

\section{Experimental Method}

\subsection{Dataset}

The STARSS23 dataset consists of real-world, multi-room recordings with annotations of event activity, spatial trajectories, and distances~\cite{shimada2024starss23}. The stereo version of STARSS23 comprises of 30,000 five-second audio recordings, with the training set containing $22.5\mathrm{h}$ of audio data. To enrich the number of real-world directional examples, we apply ACS to the STARSS23 dataset to double the amount of real data to roughly $45\mathrm{h}$.

Manual annotation of 3D SELD data is costly, resulting in the class distribution in the STARSS23 dataset to be severely imbalanced. We mitigate this challenge by first generating additional FOA data using the SpatialScaper generator~\cite{Roman2024_SpatialScaper}, before converting them into stereo audio using the provided conversion generator. In total, a total of 30,000 additional five-second synthetic stereo audio samples were generated. The combined dataset used for training therefore spans approximately $86.7\mathrm{h}$.

For feature extraction, we use a sampling rate of 24kHz, using a 1024-point FFT with the Hann window of length 1024 samples and a hop length of 300 samples, resulting in 400 time frames per audio clip. All features are mapped onto 96 Mel bands.

\subsection{Training}

The base feature stack consists of LR log-Mel spectrograms. We extend this by including the proposed perceptually-motivated features. In particular, the addition of MS log-Mel spectrograms and IV gives the \textit{MSI} feature set. The subsequent addition of MSC yields the richer \textit{MSIC} feature set. 

Spatial diversity during training is further enhanced by two alternative spectrogram-level augmentation pipelines -- either using the proposed stereo-based TFM method (ITFM) or frequency-domain perturbation that combines FilterAugment and Frequency Shifting (FAFS).  

We train each of our systems for 100 epochs using the Adam optimizer, with a peak learning rate of $1\times10^{-3}$, weight decay of $1\times10^{-4}$, and a batch size of 64. We follow the implementation of the baseline system and evaluate the model on the test split of the development set of the STARSS23 dataset, saving the model with the best validation location-dependent F-Score as our final model. Furthermore, we further fine-tune the models for another 20 epochs on only real audio recordings after the initial training. 

\section{Results}

\begin{table}[t]
\caption{3D SELD performance of the SELDNet baseline system when distance normalization is applied.}
    \centering
    \begin{tabularx}{\columnwidth}{l CCCC}
    \toprule
         Experiment
         & \Fone $\uparrow$
         & \LE $\downarrow$
         & \RDE $\downarrow$
         & \ESELD $\downarrow$ \\

         \midrule

         \bfseries Baseline & 
         23.72 & $20.8^\circ$ & 0.347 & 0.409 \\

         + DN &
         24.60 & $17.0^\circ$ & 0.287 & 0.379 \\



         \bottomrule
    \end{tabularx}
    
    \label{tab:baseline_generic_methods}
\end{table}

We employ the same validation metrics as used in DCASE 2025 Challenge Task 3. These include the location-dependent F-score (\Fone), class-dependent localization error (\LE), and class-dependent relative distance error (\RDE). In addition, we also calculate an aggregated SELD error ($\mathcal{E}_\mathrm{SELD}$) to provide an overview of the overall performance of the system as follows~\cite{politis2020overview}:

\begin{equation}
    \mathcal{E}_\mathrm{SELD} = 
    ((1 - \mathrm{F}_{\leq 20^\circ / 1}) + \frac{\mathrm{LE}_\mathrm{CD}}{180^\circ} + \mathrm{RDE}_\mathrm{CD})/3.
\end{equation}

First, we demonstrate the effectiveness of using distance normalization (DN). Table~\ref{tab:baseline_generic_methods} showcases the performance of the baseline SELDNet trained using the stereo version of the STARSS23 dataset. From the results, we can see that using DN yields consistent performance improvements across all metrics. In particular, $\mathcal{E}_\mathrm{SELD}$ decreases by $7.33\%$, showing the benefits of distance normalization in improving overall 3D SELD performance.

For our submitted systems, we use a combination of improved feature sets with different augmentation methods. Table~\ref{tab:submitted_systems} showcases the 3D SELD performance of our submitted systems. All submitted systems apply DN to the ground truth labels. 

Compared to the baseline system in Table~\ref{tab:baseline_generic_methods}, we can see that our proposed approach yields significant improvements in 3D SELD performance. We leave the optimal combination of spatial features and augmentation pipelines for future work. We theorize also that our methods, in combination with more complex and sophisticated model architectures, can yield even better performance. 

\begin{table}[t]
\caption{Performance of our submitted systems using different combinations of input features and data augmentation pipelines.}
    \centering
    \begin{tabularx}{\columnwidth}{Ll cccc}
    \toprule
         & Setup
         & \Fone $\uparrow$
         & \LE $\downarrow$
         & \RDE $\downarrow$
         & \ESELD $\downarrow$ \\

         \midrule

         \textbf{A} & MSI + ITFM &
         43.95 & $13.2^\circ$ & 0.271 & 0.302 \\

         \textbf{B} & MSIC + ITFM &
         43.12 & $\textbf{12.7}^\circ$ & \textbf{0.259} & 0.300 \\

         \textbf{C} & MSI + FAFS &
         \textbf{45.32} & $13.2^\circ$ & 0.262 & \textbf{0.294} \\
         
         \textbf{D} & MSIC + FAFS &
         43.44 & $13.2^\circ $& 0.261 & 0.300 \\

         \bottomrule
    \end{tabularx}
    
    \label{tab:submitted_systems}
\end{table}

\section{Conclusion}

This technical report details our proposed methods for the stereo-based 3D SELD task. We use perceptually-motivated input features to improve both localization and distance estimation performance. We introduce FilterAugment for the 3D SELD task, and propose a stereo-specific form of spectrogram masking augmentation. Overall, our proposed approach yields consistent and extensive improvements across all 3D SELD metrics, and can be applied to generic stereo-based 3D SELD methodologies.

\bibliographystyle{IEEEtran}
\bibliography{refs}

\end{sloppy}
\end{document}